\documentclass[prb,superscriptaddress,showpacs,floatfix,tightenlines,10pt,twocolumn]{revtex4}
\usepackage{amssymb}
\usepackage{amsmath}
\usepackage{graphicx}

\begin{document}

\title{RD-NMR\ spectra of the crystal states of the two-dimensional electron
gas in a quantizing magnetic field}
\author{R. C\^{o}t\'{e}}
\affiliation{D\'{e}partement de physique, Universit\'{e} de Sherbrooke, Sherbrooke, Qu%
\'{e}bec, J1K 2R1, Canada}
\author{Alexandre M. Simoneau}
\affiliation{D\'{e}partement de physique, Universit\'{e} de Sherbrooke, Sherbrooke, Qu%
\'{e}bec, J1K 2R1, Canada}
\keywords{Resistively-detected NMR, Wigner crystal,Skyrme crystal,quantum
Hall stripes }
\pacs{73.43.-f,73.21.Fg,73.20.Qt }

\begin{abstract}
Transport experiments on the two-dimensional electron gas (2DEG) confined
into a semiconductor quantum well and subjected to a quantizing magnetic
field have uncovered a rich variety of uniform and nonuniform phases such as
the Laughlin liquids, the Wigner, bubble and Skyrme crystals and the quantum
Hall stripe state. Optically pumped nuclear magnetic resonance (OP-NMR) has
also been extremely useful in studying the magnetization and dynamics of
electron solids with exotic spin textures such as the Skyrme crystal.
Recently, it has been demonstrated that a related technique,
resistively-detected nuclear magnetic resonance (RD-NMR), could be a good
tool to study the topography of the electron solids in the fractional and
integer quantum Hall regimes. In this work, we compute theoretically the
RD-NMR line shapes of various crystal phases of the 2DEG and study the
relation between their spin density and texture and their NMR spectra. This
allows us to evaluate the ability of the RD-NMR to discriminate between the
various types of crystal states.
\end{abstract}

\date{\today }
\maketitle

\section{INTRODUCTION}

A two-dimensional electron gas (2DEG) confined into a semiconductor quantum
well and subjected to a perpendicular magnetic field $\mathbf{B}$
experiences a great diversity of ground states as the filling factor $\nu
=N_{e}/N_{\varphi }$ is varied. Here $N_{e}$ is the number of electrons and $%
N_{\varphi }=S/2\pi \ell ^{2}$ is the Landau level degeneracy with $S$ the
2DEG area and $\ell =\sqrt{\hslash c/eB}$ the magnetic length. Besides the
well known (uniform) Laughlin states responsible for the fractional and
integer quantum Hall effects\cite{ReviewQHE}, many nonuniform phases are
also possible. These states have spatial modulations of the electronic
and/or spin densities. They have been the subject of intense theoretical and
experimental work using a variety of techniques over the past 35 years.
Famous examples are the Wigner crystal,\cite{Wigner,Wignerrevue} the bubble
crystals, the quantum Hall stripe state\cite%
{Revuebulles,Microwave,Yoshioka,Goerbig,Cotebubble,Cotestripes} and the
Skyrme crystal.\cite{Sondhi,Revueskyrmion}

Experimental evidence for the crystal phases can be obtained from microwave
absorption experiments\cite{Microwave} which detect the gapped phonon mode
of the crystal pinned by disorder. In the case of the Skyrme crystal which
has an exotic spin texture, optically pumped nuclear magnetic resonance
(OP-NMR)\cite{Barrett} has been particularly useful in measuring the
magnetization of the crystal as a function of the filling factor and Zeeman
coupling as well as its low-lying collective excitations which involve two
gapless (Goldstone) modes.\cite{Skyrmecrystal} In OP-NMR, the magnetization
is obtained by measuring the Knight shift ($K_{s}$) of the $^{71}$Ga nuclei
around $\nu =1$ in GaAs/AlGaAs multiple quantum wells. This technique
exploits the Fermi contact hyperfine interaction that exists between the
magnetic moment of the $^{71}$Ga nuclei in the GaAs quantum wells and the
spin of the electrons in the confined 2DEG. A nonzero local spin
polarization of the electrons modifies the local magnetic field seen by the
nuclei and shift their resonance frequency by an amount $K_{S}$ which is
proportional to the electronic spin polarization.

In a related technique: resistively detected nuclear magnetic resonance\cite%
{Desrat} (RD-NMR), the Knight shift of the $^{75}$As nuclei in the quantum
well is obtained from the change in the longitudinal resistance $R_{xx}$ of
the 2DEG. This change, $\Delta R_{xx},$ is related to the change in the
electronic Zeeman energy which is caused by the coupling via the hyperfine
interaction of the average nuclear magnetic moment $\left\langle
I_{Z}\right\rangle $ and the electron Zeeman energy $E_{z}\propto \left(
B+b_{0}\left\langle I_{Z}\right\rangle \right) ,$ where $B$ is the external
magnetic field and $b_{0}$ is a constant. This technique has been used to
study the liquid-solid phase transition of the 2DEG at small filling factor
in Landau levels $N=0$ and $N=1.$\cite{Gervais,Tiemann} It has also been
applied to the Skyrme crystal (or skyrmion liquid) near $\nu =1$ where an
anomalous spectral line shape is detected.\cite%
{Desrat,Gervais,Gervais2,Tracy,Barrett}

The RD-NMR technique has been used recently\cite{Tiemann52,Tiemann,Tiemann2}
to study the spatial modulations of the electronic density (or, more
precisely, of the spin density) i.e. the topography of the electron solids
in the integer and quantum Hall regimes. It is well known (see for example
Ref. \onlinecite{Cotebubble}) that the density pattern of a Wigner crystal
changes with the Landau level index $N$ because the wave function of an
electron in a magnetic field depends on $N$. This change in topography is
reflected to some degree in the NMR spectral line shape. Indeed, in
experiments, the Wigner crystals detected at small filling factors in Landau
levels $N=0$ and $N=1$ show very different RD-NMR spectra.\cite{Tiemann} In
the present work, we want to find out how different the spectral line shape
of other crystal states (bubble, stripe, and Skyrme crystals) are from one
another and how well the RD-NMR\ technique can discriminate between them. Is
the RD-NMR\ technique sufficiently sensitive, for exemple, to distinguish
between bubble crystals with $2$ or $3$ electrons per site or to separate a
bubble crystal from a stripe state? To study this sensitivity, we compare
the RD-NMR spectra of several crystal phases where the quasiparticles that
crystallize can be electrons, bubbles, skyrmions or the corresponding
anti-quasiparticles.

The calculation of the spectral line shape requires a knowledge of the
average local spin polarization $\left\langle S_{z}\left( \mathbf{r}%
,z\right) \right\rangle $ along the direction of the applied magnetic field.
Here $\mathbf{r}$ is a vector in the plane of the 2DEG and $z$ is in the
perpendicular (confining) direction. To compute $\left\langle S_{z}\left( 
\mathbf{r},z\right) \right\rangle $, we use an equation of motion method
developed previously for the study of the Wigner crystal.\cite{Cotemethode}
In this method, the spin polarization is computed from the Hartree-Fock
equation of motion of the single-particle Green's function. This method has
the advantage over the Maki-Zotos wave function\cite{Maki} that it can
easily be adapted to study crystal structures with ferromagnetic or
antiferromagnetic order,\cite{Wignerspin} crystals with more complex spin
textures such as meron or skyrmion crystals,\cite{Skyrmecrystal} or bubble
and stripe crystals.\cite{Cotebubble,Cotestripes,Ettouhami} A limitation of
this approach, however, is that the Hartree-Fock approximation does not take
into account the quantum or thermal fluctuations of the crystals nor the
disorder. As previously discussed in the NMR litterature,\cite%
{Desrat,Gervais2,Tracy} these effects may alter significantly the NMR\
spectra. However, by introducing a Gaussian blurring factor\cite{Tiemann} in
the calculation of the crystal polarization, it is possible to model the
effect of these fluctuations on the spectral line shape. In this work, we
limit ourselves to computing the RD-NMR spectra at zero temperature i.e. in
the ideal case of frozen crystals. In this regime, the nuclei in the quantum
well see an effective magnetic field that depends on their location and so
their Knight shift varies spatially. The RD-NMR\ signal is obtained by
summing the Knight-shifted signal of all nuclei as discussed in the next
section.

Our paper is organized in the following way. In Secs. II and III, we give a
brief summary of the calculation of the RD-NMR spectral line shape for
crystal states and of the spin polarization $\left\langle S_{z}\left( 
\mathbf{r},z\right) \right\rangle $ in the Hartree-Fock approximation. In
Sec. IV, we describe the different crystal phases that we include in our
study. In Sec. V, we compute the RD-NMR spectra of these crystals and
discuss how well the RD-NMR technique allows to discriminate between them.
We conclude in Sec. VI.

\section{CALCULATION OF THE RD-NMR LINE SHAPES}

We consider a 2DEG confined in a GaAs/AlGaAs quantum well. The 2DEG is in a
perpendicular magnetic field $\mathbf{B}$ that quantizes the kinetic energy
into Landau levels. We define $n_{e}$ as the electronic density in the last
partially filled Landau level $N$. The filling factor of that level is $\nu
=2\pi n_{e}\ell ^{2}$. For convenience, we define the dimensionless density $%
\widetilde{n}_{\alpha }\left( \mathbf{r}\right) $ of each spin specie $%
\alpha =\uparrow ,\downarrow $ in the partially filled Landau level $N$ by%
\begin{equation}
\widetilde{n}_{\alpha }\left( \mathbf{r}\right) =\frac{1}{2\pi \ell ^{2}}%
n_{\alpha }\left( \mathbf{r}\right) ,
\end{equation}%
where $n_{\alpha }\left( \mathbf{r}\right) $ is the density of electrons
with spin $\alpha $. The averaged spin polarization density in the direction
of the applied magnetic field is then%
\begin{equation}
\left\langle S_{z}\left( \mathbf{r,}z\right) \right\rangle =\frac{\hslash }{%
4\pi \ell ^{2}}m_{z}\left( \mathbf{r}\right) \left\vert \psi _{\nu }\left(
z\right) \right\vert ^{2},  \label{deux}
\end{equation}%
where%
\begin{equation}
m_{z}\left( \mathbf{r}\right) =\left\langle \widetilde{n}_{\uparrow }\left( 
\mathbf{r}\right) \right\rangle -\left\langle \widetilde{n}_{\downarrow
}\left( \mathbf{r}\right) \right\rangle ,
\end{equation}%
and $\psi _{\nu }\left( z\right) $ is the wave function of the first bound
state of the quantum well which is the only occupied state at sufficiently
low temperature. This wave function depends on the exact shape of the
quantum well which in turn depends on the potential of the gate used to
control the electron density. The notation $\left\langle S_{z}\left( \mathbf{%
r,}z\right) \right\rangle $ stands for the quantum and thermal averages of
the spin density. In this paper, we work at zero temperature and so $%
\left\langle S_{z}\left( \mathbf{r,}z\right) \right\rangle $ is the ground
state average of the spin density.

Following Tiemann \textit{et al}. (see Supplemental material of Ref. %
\onlinecite{Tiemann}), we write the spectral function for the RD-NMR\ signal
from the $^{75}$As nuclei in the quantum well as%
\begin{eqnarray}
I_{\nu }\left( f-f_{0}\right) &=&\frac{A}{WS}\int_{-W/2}^{+W/2}dz\left\vert
\psi _{\text{read}}\left( z\right) \right\vert ^{2}  \label{une} \\
&&\times \int_{S}d\mathbf{r}g\left( f-f_{0}+\beta \left\langle S_{z}\left( 
\mathbf{r},z\right) \right\rangle \right) ,  \notag
\end{eqnarray}%
where $A$ is a constant and $W$ is the width of the quantum well. In Eqs. (%
\ref{une}), $\psi _{\text{read}}\left( z\right) $ is the subband wave
function at filling factor $\nu _{\text{read}}$ used to account for the
position dependence of the readout sensitivity and $\beta $ is a parameter
to be determined experimentally. The function $Ag\left( f-f_{0}\right) $ is
the intrinsic line shape for the $^{75}$As nuclei in the crystal environment
(in the absence of the 2DEG). It is assumed to have the Gaussian form 
\begin{equation}
g\left( f-f_{0}\right) =e^{-\left( f-f_{0}\right) ^{2}/\Gamma ^{2}},
\end{equation}%
where $\Gamma $ is the spectral line shape and $f_{0}$ the resonance
frequency. In Eq. (\ref{une}), it is implicitly assumed that the subband
wave function is negligible outside the quantum well. This wave function
leads to an asymmetrical broadening of the spectral line shape. The
variations of the electronic density in the plane of the well introduce a
supplementary inhomogeneous broadening of the line shape.

In our numerical calculation, we compute the wave functions $\psi _{\nu
}(z_{i})$ and $\psi _{\text{read}}\left( z_{i}\right) $ in the well at $%
n_{d} $ points $z_{i}$ and $m_{z}\left( \mathbf{r}_{j}\right) $ at $n_{t}$
points $\mathbf{r}_{j}$ in the unit cell of the crystal. Upon defining 
\begin{equation}
\overline{\psi }_{\nu }(z_{i})=\sqrt{\frac{W}{n_{d}}}\psi _{\nu }(z_{i}),
\end{equation}%
(and a similar relation for $\overline{\psi }_{\text{read}}(z_{i})$) which
satisfies 
\begin{equation}
\sum_{i=1}^{n_{d}}\left\vert \overline{\psi }_{\nu }(z_{i})\right\vert
^{2}\approx 1,
\end{equation}%
we get%
\begin{eqnarray}
I_{\nu }\left( f-f_{0}\right) &=&\frac{A}{Wn_{t}}\sum_{i=1}^{n_{d}}%
\sum_{j=1}^{n_{t}}\left\vert \overline{\psi }_{read}(z_{i})\right\vert ^{2}
\label{trois} \\
&&\times g\left( f-f_{0}+\zeta n_{d}m_{z}\left( \mathbf{r}_{j}\right)
\left\vert \overline{\psi }_{\nu }\left( z_{i}\right) \right\vert
^{2}\right) ,  \notag
\end{eqnarray}%
where $\zeta $ is now the constant to be found experimentally.

Since $n_{t}$ can be large in Eq. (\ref{trois}), $I_{\nu }\left(
f-f_{0}\right) $ is easier to evaluate if we first compute a function giving
the probability of occurrence of $m_{z}$. We find the minimal and maximal
values of $m_{z}$ in a unit cell, divide the interval between these two
values into $n_{b}$ bins of width $\Delta m_{z}$ with middle values $m_{z,k}$
and write $P\left( m_{z,k}\right) $ for the probability of occurrence of $%
m_{z,k}$. The signal then becomes%
\begin{eqnarray}
I_{\nu }\left( f\right) &=&\frac{A}{W}\sum_{i=1}^{n_{d}}\sum_{k=1}^{n_{b}}P%
\left( m_{z,k}\right) \left\vert \overline{\psi }_{read}(z_{i})\right\vert
^{2}  \label{signal2} \\
&&\times g\left( f-f_{0}+\zeta n_{d}m_{z,k}\left\vert \overline{\psi }_{\nu
}\left( z_{i}\right) \right\vert ^{2}\right) ,  \notag
\end{eqnarray}%
with%
\begin{equation}
\sum_{k=1}^{n_{b}}P_{k}=1.
\end{equation}%
For a uniform state with $m_{z}\left( \mathbf{r}\right) =m_{z},$%
\begin{eqnarray}
I_{\nu ,\text{uniform}}\left( f\right) &=&\frac{A}{W}\sum_{i=1}^{n_{d}}\left%
\vert \overline{\psi }_{read}(z_{i})\right\vert ^{2}  \label{uniform} \\
&&\times g\left( f-f_{0}+\zeta n_{d}m_{z}\left\vert \overline{\psi }_{\nu
}\left( z_{i}\right) \right\vert ^{2}\right) .  \notag
\end{eqnarray}%
Even in this simple case, the line shape is complex since it depends on the
form of the confining wave function.

In the experimental setup described in Ref. \onlinecite{Tiemann}, the
magnetic field is kept fixed. A doping layer provides a finite density of
electrons in the well. This density can be modified by changing the
potential of a gate allowing the RD-NMR signal to be studied in some range
of filling factors. The ionized doping layer, the applied electric field and
the finite density of electrons in the well all modify its potential profile
and consequently the subband wave functions $\psi _{\nu }\left( z\right) $
and $\psi _{\text{read}}\left( z\right) .$ These functions must then be
found by solving the Schr\"{o}dinger-Poisson equations. To fit the
experimental RD-NMR signal correctly, these two wave functions must be
computed from the exact potential profile of the quantum well at filling
factors $\nu $ and $\nu _{\text{read}}.$

In this work, we are interested in studying how a qualitative change in the
crystal structure (such as the transition from a Wigner to a bubble crystal
with two electrons per site) affects the RD-NMR signal. To isolate the
effect of a change in the spin polarization, we assume an ideal situation
where the quantum well profile is fixed, i.e. unchanged by a variation of
the filling factor. We thus approximate both functions $\psi _{\nu }\left(
z\right) $ and $\psi _{\text{read}}\left( z\right) $ by the simple form%
\begin{equation}
\psi \left( z\right) =\sqrt{\frac{2}{W}}\cos \left( \frac{\pi z}{W}\right)
\label{onde}
\end{equation}%
which is the first bound state of an infinite quantum well. Our final
expression for the RD-NMR spectral line shape is thus 
\begin{eqnarray}
I\left( f\right) &=&\frac{2A}{n_{d}}\sum_{i=1}^{n_{d}}\sum_{k=1}^{n_{b}}P%
\left( m_{z,k}\right) \cos ^{2}\left( \frac{\pi z_{i}}{W}\right)
\label{signal} \\
&&\times g\left( f-f_{0}+2\zeta m_{z,k}\cos ^{2}\left( \frac{\pi z_{i}}{W}%
\right) \right) .  \notag
\end{eqnarray}

For greater clarity in the figures on this paper, we plot $P\left(
m_{z,k}\right) $ as a probability density function $P\left( m_{z}\right) $
instead of as an histogram of values of $m_{z,k}$. We normalize $P\left(
m_{z}\right) $ such that:%
\begin{equation}
\int_{m_{z,\min }}^{m_{z,\max }}P\left( m_{z}\right) dm_{z}=1.
\end{equation}%
In our calculations, we discretize the confining wave functions into $%
n_{d}=200$ points, the unit cell of each crystal into $500\times 500$ points
($n_{t}=250000$) that are further classified into $n_{b}=200$ bins.

\section{SPIN POLARIZATION DENSITY IN THE HARTREE-FOCK APPROXIMATION}

The electronic density in the partially filled level is written as%
\begin{equation}
\widetilde{n}_{\alpha }\left( \mathbf{r}\right) =\sum_{\mathbf{G}%
}\left\langle \rho _{\alpha ,\alpha }\left( \mathbf{G}\right) \right\rangle
L_{N}^{0}\left( \frac{G^{2}\ell ^{2}}{2}\right) e^{-G^{2}\ell ^{2}/4}e^{i%
\mathbf{G}\cdot \mathbf{r}},  \label{density}
\end{equation}%
where $L_{N}^{0}\left( x\right) $ is an associated Laguerre polynomial and
the operator 
\begin{eqnarray}
\rho _{\alpha ,\beta }\left( \mathbf{G}\right) &\equiv &\frac{1}{N_{\varphi }%
}\sum_{X,X^{\prime }}e^{-\frac{i}{2}G_{x}\left( X+X^{\prime }\right) }
\label{rho} \\
&&\times \delta _{X,X^{\prime }+G_{y}\ell ^{2}}c_{N,X,\alpha }^{\dagger
}c_{N,X^{\prime },\beta }.  \notag
\end{eqnarray}%
The set of vectors $\left\{ \mathbf{G}\right\} $ are the reciprocal lattice
vectors of the crystal considered while the set of ground-state average
values $\left\{ \left\langle \rho _{\alpha ,\alpha }\left( \mathbf{G}\right)
\right\rangle \right\} $ can be considered as the order parameters of the
crystal. They are related to the single-particle Matsubara Green's function
by%
\begin{equation}
\left\langle \rho _{\beta ,\alpha }\left( \mathbf{G}\right) \right\rangle
=G_{\alpha ,\beta }\left( \mathbf{G,}\tau =0^{-}\right) ,
\end{equation}%
where 
\begin{eqnarray}
G_{\alpha ,\beta }\left( \mathbf{G,}\tau \right) &=&\frac{1}{N_{\varphi }}%
\sum_{X,X^{\prime }}e^{-\frac{i}{2}G_{x}\left( X+X^{\prime }\right) } \\
&&\times \delta _{X,X^{\prime }-G_{y}\ell ^{2}}G_{\alpha ,\beta }\left(
X,X^{\prime },\tau \right)  \notag
\end{eqnarray}%
and%
\begin{equation}
G_{\alpha ,\beta }\left( X,X^{\prime },\tau \right) =-\left\langle
Tc_{N,X,\alpha }\left( \tau \right) c_{N,X^{\prime },\beta }^{\dagger
}\left( 0\right) \right\rangle ,
\end{equation}%
with $T$ the time-ordering operator and $\tau $ the imaginary time. In Eq. (%
\ref{rho}), the operator $c_{N,X,\alpha }^{\dagger }$ creates an electron in
Landau level $N$ with guiding-center index $X$ in the Landau gauge and with
spin $\alpha $. Since we consider the filled levels below $N$ as inert (they
do not contribute to the spin polarization), we hereafter drop the index $N$
to simplify the notation

The single-particle Green's function is computed by using an equation of
motion method\cite{Cotemethode} which is generalized to include the spin
degree of freedom and the possibility of a spin texture in the plane of the
quantum well. The Hartree-Fock equation of motion for $G_{\alpha ,\beta
}\left( \mathbf{G,}i\omega _{n}\right) ,$ where $\omega _{n}$ are the
Matsubara frequencies, is%
\begin{eqnarray}
&&\left[ i\omega _{n}-\left( E_{\alpha }-\mu \right) /\hslash \right]
G_{\alpha ,\beta }\left( \mathbf{G},\omega _{n}\right) =\delta _{\mathbf{G}%
,0}\delta _{\alpha ,\beta }  \label{equation} \\
&&+\frac{1}{\hslash }\sum_{\mathbf{G}^{\prime }\neq \mathbf{G}}U^{H}\left( 
\mathbf{G-G}^{\prime }\right) e^{-i\mathbf{G}\times \mathbf{G}^{\prime }\ell
^{2}/2}G_{\alpha ,\beta }\left( \mathbf{G}^{\prime },\omega _{n}\right) 
\notag \\
&&-\frac{1}{\hslash }\sum_{\mathbf{G}^{\prime }}\sum_{\gamma }U_{\alpha
,\gamma }^{F}\left( \mathbf{G-G}^{\prime }\right) e^{-i\mathbf{G}\times 
\mathbf{G}^{\prime }\ell ^{2}/2}G_{\gamma ,\beta }\left( \mathbf{G}^{\prime
},\omega _{n}\right) ,  \notag
\end{eqnarray}%
with the potentials%
\begin{eqnarray}
U^{H}\left( \mathbf{G}\right) &=&\left( \frac{e^{2}}{\kappa \ell }\right)
H\left( \mathbf{G}\right) \left\langle \rho \left( \mathbf{G}\right)
\right\rangle , \\
U_{\alpha ,\beta }^{F}\left( \mathbf{G}\right) &=&\left( \frac{e^{2}}{\kappa
\ell }\right) X\left( \mathbf{G}\right) \left\langle \rho _{\beta ,\alpha
}\left( \mathbf{G}\right) \right\rangle ,
\end{eqnarray}%
where $\left\langle \rho \left( \mathbf{G}\right) \right\rangle \equiv
\sum_{\alpha }\left\langle \rho _{\alpha ,\alpha }\left( \mathbf{G}\right)
\right\rangle $ and $\kappa $ is the dielectric constant of the host
semiconductor. The non-interacting single-particle energy is given by%
\begin{equation}
E_{\alpha }=-\alpha \frac{\left\vert g^{\ast }\right\vert \mu _{B}B}{2},
\end{equation}%
where $g^{\ast }=-0.44$ is the effective $g-$factor of bulk GaAs and $\mu
_{B}$ is the Bohr magneton. The Hartree and Fock interactions are defined by%
\begin{eqnarray}
H\left( \mathbf{q}\right) &=&\left( \frac{e^{2}}{\kappa \ell }\right) \frac{1%
}{q\ell }e^{-q^{2}\ell ^{2}/2}\Lambda \left( q\ell \right) \left[
L_{N}^{0}\left( \frac{q^{2}\ell ^{2}}{2}\right) \right] ^{2}, \\
X\left( \mathbf{q}\right) &=&\left( \frac{e^{2}}{\kappa \ell }\right) \sqrt{2%
}\int_{0}^{\infty }dxe^{-x^{2}}\Lambda \left( x\right) \left[
L_{N}^{0}\left( x^{2}\right) \right] ^{2} \\
&&\times J_{0}\left( \sqrt{2}xq\ell \right) ,  \notag
\end{eqnarray}%
where $J_{0}\left( x\right) $ is the Bessel function of order zero. Because
of the Laguerre polynomial $L_{N}^{0}\left( x\right) $, these interactions
depend on the Landau level index $N.$ The function $\Lambda \left( x\right) $
takes into account the finite width of the quantum well. It is defined by 
\begin{equation}
\Lambda \left( x\right) =\int_{-\frac{W}{2}}^{+\frac{W}{2}}dz\left\vert \psi
\left( z\right) \right\vert ^{2}\int_{-\frac{W}{2}}^{+\frac{W}{2}}dz^{\prime
}\left\vert \psi \left( z^{\prime }\right) \right\vert ^{2}e^{-x\left\vert
z-z^{\prime }\right\vert /\ell }
\end{equation}%
and depends on the magnetic field $B.$ For an ideal 2DEG, $\Lambda \left(
x\right) =1.$

Equation (\ref{equation}) is self-consistent and can be solved numerically
by successive iterations.\cite{Cotemethode} A Gaussian blurring $%
e^{-r^{2}/\sigma ^{2}}$ of the density $\widetilde{n}_{\alpha }\left( 
\mathbf{r}\right) $ can be obtained by multiplying the summand in Eq. (\ref%
{density}) by the factor $e^{-\sigma ^{2}G^{2}\ell ^{2}/4}.$ This blurring
can be used to mimic the effects of quantum and thermal fluctuations on the
crystal. The parameter $\sigma $ is then adjusted, in this ase, so as to fit
the experimental spectra.\cite{Tiemann}

\section{DESCRIPTION OF THE CRYSTAL PHASES}

We give now a brief description of the different crystal phases for which we
want to compute the NMR spectral line shape. In the Hartree-Fock
approximation, even the quantum Hall stripe phase has density modulations
along the direction of the stripes and so can be described as a crystal
although a very anisotropic one (a face-centered rectangular a large aspect
ratio). All these states can by described formally as crystals of some
quasiparticle. The quasiparticle can be an electron, an electron bubble, a
skyrmion or the corresponding quasi-antiparticles:\ hole, hole bubble,
antiskyrmion. We write $n_{\text{qp}}\left( \mathbf{r}\right) $ for the
density of these quasiparticles and define $\nu \in \left[ 0,2\right] $ as
the filling factor of the partially filled level $N$ (the total filling
factor is $2N+\nu $). The filled levels are considered as inert. A level
with both spin sub-levels filled is unpolarized and so does not contribute
to the NMR\ spectra.

We consider the following phases:

\begin{itemize}
\item BC1 ($\nu \leq 0.5$). A bubble crystal, as described in Ref. %
\onlinecite{Cotebubble}, where the quasiparticle at each site is a $M-$%
electron bubble with spin up. The densities $\widetilde{n}_{\uparrow }\left( 
\mathbf{r}\right) =n_{\text{qp}}\left( \mathbf{r}\right) ,\widetilde{n}%
_{\downarrow }\left( \mathbf{r}\right) =0$ and the spin polarization density
is $m_{z}\left( \mathbf{r}\right) =n_{\text{qp}}\left( \mathbf{r}\right) .$

\item BC2 ($0.5\leq \nu \leq 1$). A bubble crystal where the quasiparticle
is a $M-$hole bubble with spin down. The densities $\widetilde{n}_{\uparrow
}\left( \mathbf{r}\right) =1-n_{\text{qp}}\left( \mathbf{r}\right) ,%
\widetilde{n}_{\downarrow }\left( \mathbf{r}\right) =0$ and $m_{z}\left( 
\mathbf{r}\right) =1-n_{\text{qp}}\left( \mathbf{r}\right) .$

\item BC3 ($1\leq \nu \leq 1.5$). A bubble crystal where the quasiparticle
is a $M$-electron bubble with spin down. The densities $\widetilde{n}%
_{\uparrow }\left( \mathbf{r}\right) =1,\widetilde{n}_{\downarrow }\left( 
\mathbf{r}\right) =n_{\text{qp}}\left( \mathbf{r}\right) $ and $m_{z}\left( 
\mathbf{r}\right) =1-n_{\text{qp}}\left( \mathbf{r}\right) .$

\item BC4 ($1.5\leq \nu \leq 2$). A bubble crystal where the quasiparticle
is a $M-$hole bubble with spin up. The densities $\widetilde{n}_{\uparrow
}\left( \mathbf{r}\right) =1,\widetilde{n}_{\downarrow }\left( \mathbf{r}%
\right) =1-n_{\text{qp}}\left( \mathbf{r}\right) $ and $m_{z}\left( \mathbf{r%
}\right) =n_{\text{qp}}\left( \mathbf{r}\right) .$
\end{itemize}

The special case $M=1$ corresponds to the Wigner crystal. The Hartree-Fock
Hamiltonian being electron-hole symmetric, $n_{\text{qp}}\left( \mathbf{r}%
\right) $ is the same function for all four phases. It follows that BC1 and
BC4 have the same spectrum and similarly for BC2 and BC3. Moreover, the
spectrum for BC2 (or BC4) is obtained from the polarizations $m_{z,k}$
computed for BC1 (or BC3) by replacing $m_{z,k}$ in Eq. (\ref{signal}) by $%
1-m_{z,k}$.

The description given above for BC1 to BC4 applies with no change to the
quantum Hall stripe phases when considered as face-centered rectangular
Wigner crystal ($M=1$). In this case, the aspect ratio that minimizes the
ground-state energy depends on the Landau level index $N$ and filling factor 
$\nu $. To distinguish the stripe phases from the bubble crystals, we use
for the former the abbreviations SC1 to SC4.

The last phases we consider are the Skyrme crystals. These occur for $\nu $
close to $1$:

\begin{itemize}
\item SK1 ($\nu <1$). A crystal where the quasiparticle at each site is an
anti-skyrmion of topological charge $Q=-1.$ The densities can be described
by $\widetilde{n}_{\uparrow }\left( \mathbf{r}\right) =1-n_{a}\left( \mathbf{%
r}\right) ,\widetilde{n}_{\downarrow }\left( \mathbf{r}\right) =n_{b}\left( 
\mathbf{r}\right) .$ They depend on the ratio of the Zeeman coupling to the
Coulomb energy $\gamma =\left\vert g^{\ast }\right\vert \mu _{B}B/\left(
e^{2}/\kappa \ell \right) $ and on the filling factor$.$ The density $%
m_{z}\left( \mathbf{r}\right) =1-n_{a}\left( \mathbf{r}\right) -n_{b}\left( 
\mathbf{r}\right) $ while $\widetilde{n}\left( \mathbf{r}\right)
=1-n_{a}\left( \mathbf{r}\right) +n_{b}\left( \mathbf{r}\right) =1-n_{\text{%
qp}}\left( \mathbf{r}\right) .$

\item SK2 ($\nu >1$). A crystal where the quasiparticle at each site is a
skyrmion of topological charge $Q=1.$ The densities $\widetilde{n}_{\uparrow
}\left( \mathbf{r}\right) =1-n_{b}\left( \mathbf{r}\right) ,\widetilde{n}%
_{\downarrow }\left( \mathbf{r}\right) =n_{a}\left( \mathbf{r}\right) $ and
so $m_{z}\left( \mathbf{r}\right) =1-n_{a}\left( \mathbf{r}\right)
-n_{b}\left( \mathbf{r}\right) $ and $\widetilde{n}\left( \mathbf{r}\right)
=1+n_{\text{qp}}\left( \mathbf{r}\right) .$
\end{itemize}

By electron-hole symmetry, SK1 and SK2 have the same NMR\ spectrum.\cite%
{Note1} In the Hartree-Fock calculation,\cite{Skyrmecrystal} the
lowest-energy lattice structure for the Skyrme crystal if $\nu $ is not too
close to $1$ is a square lattice with two skyrmions of opposite phases per
unit cell i.e. the square lattice antiferromagnetic (SLA) phase. It has
lower energy than a triangular lattice where all skyrmions have the same
phase (TLF). For $\nu $ close to one, a three-sublattice crystal where
skyrmions on different sublattice are rotated by $120^{\circ }$ should be
the ground state. However, it is difficult to stabilize this phase
numerically so that we will only consider the SLA and TLF crystals in our
calculation. We remark that, of all the crystal states considered here, SK1
and SK2 are the only ones where $m_{z}\left( \mathbf{r}\right) $ can be
locally negative. As we will show, this makes the NMR\ Skyrme crystal's
spectrum quite different from that of the other crystal states.

\section{NMR SPECTRA OF THE\ CRYSTAL STATES}

In this section, we compute the RD-NMR\ spectra of the crystal states
described in Sec. IV. Because of the electron-hole symmetry, we can restrict
ourselves to the BC1,BC2,SC1,SC2 and SK1 phases. The order parameters for
the crystals are calculated for a quantum well of width $W=27$ nm and in a
magnetic field $B=6.4$ T as in the experiments.\cite{Tiemann} We work,
however, at zero temperature. We define the Knight shift as the displacement
of the peak of the spectrum from the bare resonance frequency $f_{0}.$ At $%
B=6.4$ T, the $^{75}$As nucleus has $f_{0}=46.3997$ MHz with a linewidth $%
\Gamma =1.5$ kHz\cite{Tiemann}. We use this value of $\Gamma $ in our
calculations. Figure 4(b) of Ref. \onlinecite{Tiemann} gives the expected
behavior of the Knight shift of a fully polarized uniform system for filling
factors $\nu \in \left[ 0,1/3\right] $ in Landau level $N=0$. We use this
figure to get the constant $\zeta $ that enters Eq. (\ref{signal}). We get $%
\zeta =16$. Since $\zeta >0,$ a positive spin polarization leads to a
negative Knight shift in Eq. (\ref{signal}).

\begin{figure}[tbph]
\includegraphics[scale=0.8]{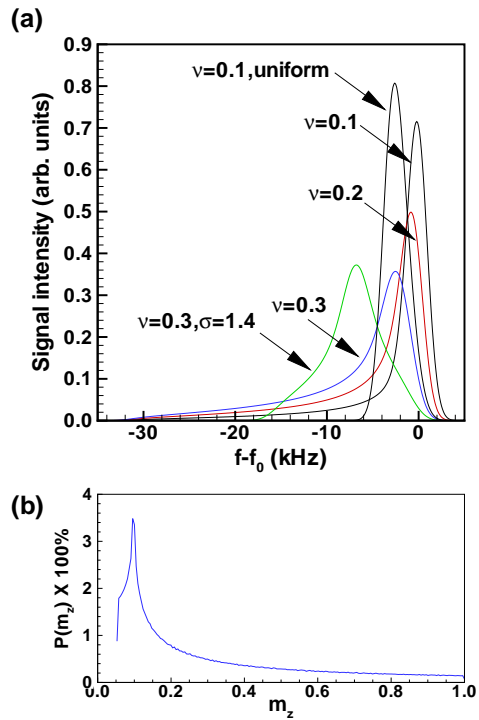}
\caption{(Color online) (a) RD-NMR spectra for the Wigner crystal phase
(BC1) at different filling factors $\protect\nu $ in Landau level $N=0$ with 
$\protect\sigma =0$ and for $\protect\nu =0.3$ with $\protect\sigma =1.4;$%
(b) Probability density $P\left( m_{z}\right) $ in the Wigner crystal phase
for $\protect\nu =0.3$, $\protect\sigma =0.$ }
\end{figure}

\subsection{Phase BC1 with $M=1$ in $N=0$}

Figure 1(a) shows the spectra for the BC1 phase with $M=1$ i.e. a Wigner
crystal at filling factors $\nu =0.1,0.2,0.3$ in Landau level $N=0.$ These
spectra were discussed previously\cite{Tiemann} but we reconsider them here
for completeness. The spectra are characterized by a single peak and a long
tail, absent in the the uniform state, which is due to the large spread in
the values of $m_{z}$ in the crystal as shown by the plot of $P\left(
m_{z}\right) $ in Fig. 1(b). From Eq. (\ref{signal}), a region of space with
a certain $m_{z}$ produces a local shift in frequency of $\approx
2m_{z}\zeta .$ Since $\zeta =16,$ this shift is $32$ kHz for the largest
possible value $m_{z}=1$. The function $P\left( m_{z}\right) $ is dominated
by the regions of small values of $m_{z}$ so that the Knight shifts in the
spectra are small. As $\nu $ increases, the most probable value of $m_{z}$
increases (not shown in the figure) and the Knight shift becomes more
negative. At the same time, the maximal intensity of the signal decreases.
The most probable value of $m_{z}$ at these filling factors is smaller that
the spatially averaged value of $m_{z}$ so that the Knight shift of the
crystal is less negative\ than that of the uniform state$.$ All spectra
start at $f-f_{0}=0$ even though the crystal and the liquid have no region
with $m_{z}=0.$ This is because the subband wave function term $\overline{%
\psi }_{\nu }\left( z_{i}\right) $ in Eq. (\ref{signal}) goes to zero at $%
z=\pm W/2.$

At $\nu =0.3,$ there is an important difference between the theoretical and
experimental spectra.\cite{Tiemann} This is attributed to the fact that the
Hartree-Fock approximation for $m_{z}\left( \mathbf{r}\right) $ neglects
both quantum and thermal fluctuations. To get a spectra closer to the
experimental result, it is necessary to use a finite blurring factor of
order $\sigma \approx 1.4$ at $\nu =0.3.$\cite{Tiemann} As shown in Fig.
1(a), this shifts the peaks to a more negative value and cuts the
low-frequency tail.

\begin{figure}[tbph]
\includegraphics[scale=0.8]{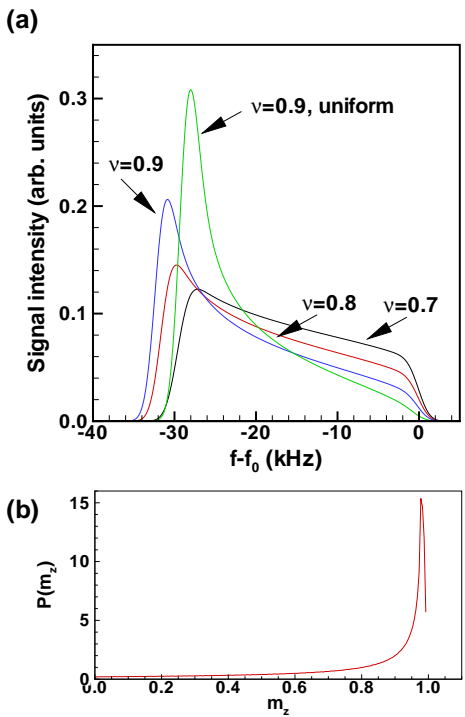}
\caption{(Color online) (a) RD-NMR spectra for the Wigner crystal phase
(BC2) at different filling factors $\protect\nu $ in Landau level $N=0$ with 
$\protect\sigma =0;$(b) Probability density $P\left( m_{z}\right) $ in the
Wigner crystal phase for $\protect\nu =0.8,\protect\sigma =0.$ }
\end{figure}

\subsection{Phase BC2 with $M=1$ in $N=0$}

Figure 2(a) shows the spectra for the BC2 phase with $M=1$ at filling
factors $\nu =0.7,0.8,0.9$ in Landau level $N=0.$ A comparison with Fig.
1(a) shows that the spectra for BC2 are distinctively different from those
of BC1. In the plot of $P\left( m_{z}\right) $ in Fig. 2(b), the most
probable value for $m_{z}$ is now close to $m_{z}=1$ and so the Knight shift
is large. This is also true for the uniform phase. As in BC1, the spread in $%
m_{z}$ is important and the signal is intense in a large range of frequency
shift$.$ In contrast to BC1, however, the maximal $m_{z}$ decreases with
decreasing $\nu $ and so does the spread of the signal. The Knight shift
becomes less negative with decreasing filling factor because the maximum
value of $m_{z}$ decreases with decreasing $\nu $. The most probable value
of $m_{z}$ in the BC2 phase is, in contrast to BC1, bigger than that of the
uniform state so that the Knight shift in the crystal BC2 state is more
negative than in the uniform state.

\begin{figure}[tbph]
\includegraphics[scale=0.8]{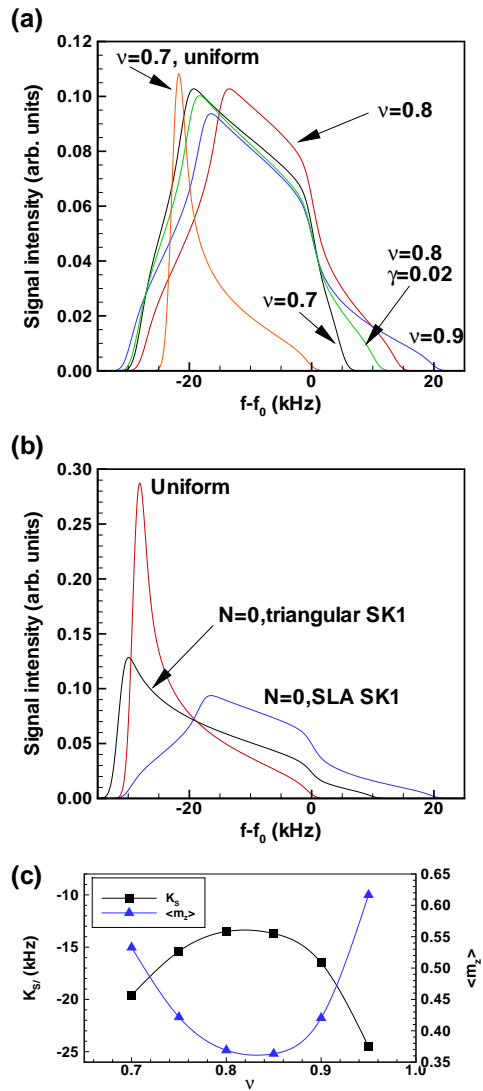}
\caption{(Color online) RD-NMR spectra in Landau level $N=0$ for: (a) the
Skyrme crystal phases (SK1) at different filling factors $\protect\nu $ and
for a Zeeman coupling $\protect\gamma =0.015$ (unless indicated otherwise)
with $\protect\sigma =0;$ (b) the triangular (TLF) and square (SLA) Skyrme
crystals for $\protect\gamma =0.015,\protect\nu =0.9$; (c) Knight shift
(measured at the peak position) and spatially averaged polarization $%
\left\langle m_{z}\right\rangle $ as a function of the filling factor.}
\end{figure}

\subsection{Phase SK1 in $N=0$}

Figure 3(a) shows the spectra of the SLA\ Skyrme crystal SK1 at filling
factors $\nu =0.7,0.8,0.9$ in Landau level $N=0$ for a Zeeman coupling $%
\gamma =g^{\ast }\mu _{B}B/\left( e^{2}/\kappa \ell \right) =0.015$ and at $%
\nu =0.8$ for $\gamma =0.02.$ For comparison, we include the spectrum of the
uniform state at $\nu =0.7$ for $\gamma =0.015.$ The polarization $%
S_{z}=\left( \nu _{\uparrow }-\nu _{\downarrow }\right) /\nu $ is shown as a
function of $\nu $ and $\gamma $ in Fig. 1 of Ref. \onlinecite{Skyrmecrystal}%
. (In that calculation, the well width was taken to be zero but taking $W=27$
nm does not change much that result.) That Hartree-Fock calculation of $%
S_{z} $ gives a fairly accurate description of the magnetization of the
Skyrme crystal observed in NMR\ experiments.\cite{Barrett} As Fig. 3(a)
shows, the spectrum for the Skyrme crystal is different from that of the
Wigner crystal [compare Fig. 2(a) and Fig. 3(a)] or the uniform state. For
the Skyrme crystal, there is large frequency range of positive Knight shift
which is due to the spatial regions of reversed spin (negative $m_{z}$) in
the skyrmion's texture. In the Wigner crystal, there is no spin texture and $%
m_{z}\left( \mathbf{r}\right) $ is always positive. For $\nu \rightarrow 1,$
the skyrmion's size and so the number of reversed spins increases meaning
that the spatial regions of positive Knight shift increase too. The regions
of positive shift decrease with increasing Zeeman coupling $\gamma $ since
that coupling decreases the skyrmion's size. Another important difference
between the spectrum of the Wigner and the Skyrme crystals is that the
former is independent of $\gamma .$

We compare the spectra of different Skyrme crystal lattices at $\nu =0.9$ in
Fig. 3(b). The SLA\ and triangular Skyrme crystals in $N=0$ are evaluated at 
$\gamma =0.015.$ They have a spatially averaged polarization $m_{z}=0.42$
and $m_{z}=0.77$ respectively. The Knight shift is larger for the triangular
lattice because of the regions of higher polarization but the positive
Knight shift extends to higher frequencies for the SLA structure which has
more reversed spins. The Skyrme crystal for a perfect 2DEG is not expected
to be the ground state for $N>0.$ Indeed, in $N=1,2$ we were not able to
stabilize a Skyrme crystal even for a Zeeman coupling as small as $\gamma
=0.0001.$ With a large number of iterations, the solution converges towards
a triangular Wigner crystal of holes (BC2) with no spin texture [its
spectrum is shown in Fig. 4(b)]. This is consistent with a recent experiment%
\cite{Tiemann2} where no Skyrme crystal are detected in $N=1$ around $\nu =1 
$ a finite temperature. The low-temperature phase of the 2DEG in this case
seems to be a Wigner crystal of holes (type BC2 with $M=1$). The difference
in the line shapes of BC2 crystals for $N=0$ and $N=1$ [compare Fig. 2(a)
with Fig. 4(b) arises because of the difference in the electronic wave
functions of the two Landau levels as we explain in more detail below.

Figure 3(c) shows the Knight shift measured at the position of the peak in
the spectrum of Fig. 3(a) and the spatially averaged polarization $m_{z}$ as
a function of the filling factor $\nu .$ For $\nu <1,$ the uniform state has 
$m_{z}=1.$ As can be seen by comparing the two curves, the relation between
the Knight shift measured in this way and the average magnetization is not
exactly linear (as is the case for zero-width uniform 2DEG).

The spectra that we gave in this section apply to an ideal situation where
the Skyrme crystals are frozen. With thermal fluctuations, the line shapes
may be very different. At finite temperature, one must know the exact regime
where the experiment is carried on i.e. the Skyrme's crystal dynamics.\cite%
{Barrett,Sinova,Villares,Gervais2} Experiments\cite{Barrett,Desrat} close to 
$\nu =1$ have measured an anomalous line shape for the Skyrme crystal that
our simple ground-state calculation cannot reproduce. in fact, it has even
been suggested that the dispersive line shape found experimentally may be
inconsistent with the usual model of RD-NMR where the signal is due solely
to a hyperfine-induced increase of the electronic Zeeman energy.\cite{Tracy}

\begin{figure}[tbph]
\includegraphics[scale=0.8]{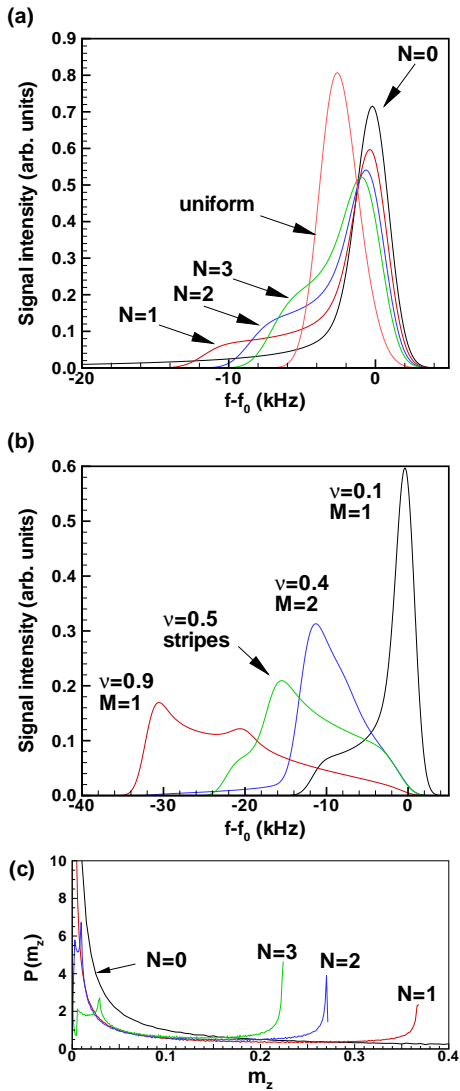}
\caption{(Color online) RD-NMR spectra for: (a) the BC1 phase with $M=1$
(Wigner crystal) at filling factor $\protect\nu =0.1$ in Landau levels $%
N=0,1,2,3$ and for the uniform phase at $\protect\nu =0.1;$ (b) the BC1, SC1
and BC2 phases in Landau level $N=1$ at different filling factors; (c)
probability density $P\left( m_{z}\right) $ for the same phases as in (a).
The function $P\left( m_{z}\right) $ for $N=0$ extends to $m_{z}=1$ (not
shown on the graph).}
\end{figure}

\subsection{Phases BC1,BC2 and SC1,SC2 in Landau levels $N>0$}

The Laguerre polynomial $L_{N}^{0}\left( x\right) $ that enters the
calculation of the density as well as the Hartree and Fock potentials causes
these functions to change with the Landau level index. In $N=0,$ the density
for a bubble crystal with $M=1$ has a Gaussian profile centered at each
lattice site. For $N>0$ and $M=1,$ the density profile has the form of a
ring centered on each lattice site. A new ring is added each time $M$
increases with the exception of the last bubble phase with $M=N+1$ in Landau
level $N.$ In this case, a peak is added at the center of the existing $M$
rings (see Fig. 2 of Ref. \onlinecite{Cotebubble}).

We compare the spectra of the $M=1$ BC1 phase for $N=0,1,2,3$ at $\nu =0.1$
in Fig. 4(a). The corresponding probability densities $P\left( m_{z}\right) $
are shown in Fig. 4(c). The most probable density is the minimal density
which increases with $N$ while its probability decreases with $N.$
Consequently, the Knight shift in Fig. 4(a) becomes more negative with $N$
but the peak intensity in the spectral line shape decreases with $N$. The
next contribution to these spectra comes from the ring. At fixed $\nu ,$ the
diameter of the ring increases with $N$ while $m_{z}$ in the ring decreases.
This behavior is reflected in the spectra shown in Fig. 4(a): the frequency
and the intensity of the shoulder in the spectrum increases with $N$ since,
as the ring expands, it covers a greater area. For comparison, we include in
Fig. 4(a) the spectrum of the uniform phase at $\nu =0.1$ which is
independent of $N.$ The peak in the density of the crystal for $N=0,M=1$
produces a long tail in its spectrum that is absent of the spectrum of a $%
M=1 $ crystal in higher Landau levels.

Figure 4(b) shows the spectra of several phases in Landau level $N=1:$ the
BC1 crystals with $M=1$ ($\nu =0.1$) and $M=2$ ($\nu =0.4$), the stripe
phase SC1 at $\nu =0.5$ and the BC2 crystal with $M=1$ at $\nu =0.9.$ The
stripe phase is not present in $N=1$ in the HFA but appears in DMRG
calculation.\cite{Yoshioka} On the contrary, the $M=2$ bubble crystal is
present in HFA\ but not in DMRG. The transition from $M=1$ to $M=2$ is
predicted to occur at $\nu =0.36$ in the HFA. The phase diagram obtained in
DMRG is richer than that of the HFA because it contains the fractional
quantum Hall states and other correlated liquid states. In contrast to the
HFA, it predicts some form of stripe state in $N=0$ (whose structure is
however different from the stripe state found in higher Landau levels) and a
stripe state in $N=1$ when $\nu \in \left[ 0.38,0.47\right] $ whose
structure is similar to the stripe state in higher Landau levels predicted
by the HFA. Up to now, only the $M=1$ BC1 phase has been detected
experimentally.\cite{Tiemann2} The long tail in the $N=1,M=2$ phase, like
that of the Wigner crystal with $N=0,M=1$ is due to the peak in density.
This makes its spectrum different from that of the other phases in $N=1$.
The spectra of the different phases in $N=1$ are quite different and their
detection should allow the unambiguous identification of the corresponding
phases.

\begin{figure}[tbph]
\includegraphics[scale=0.8]{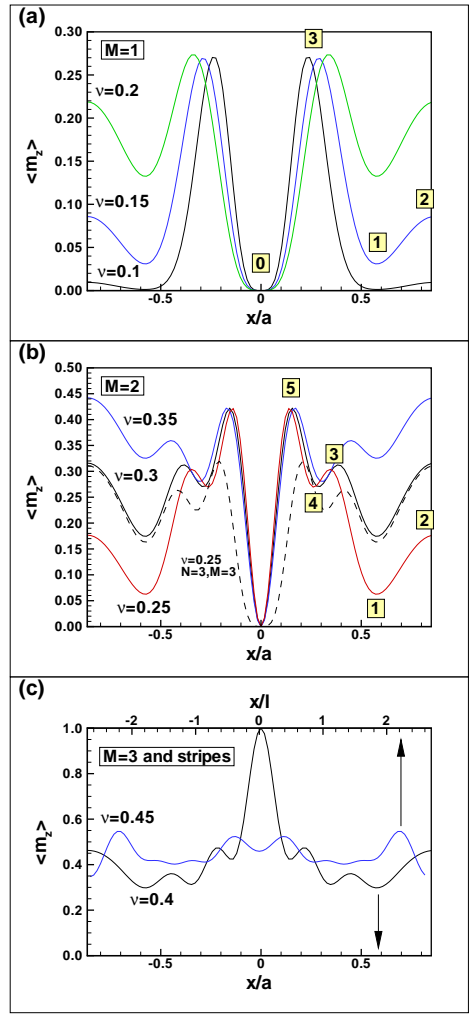}
\caption{(Color online) Traces of the spin density $m_{z}\left( \mathbf{r}%
\right) $ along $x$ for $y=0$ and at different filling factors in $N=2$ for
the : (a) $M=1$ bubble crystals; (b) $M=2$ bubble crystals; (c) $M=3$ bubble
crystals (bottom $x$ axis) and stripe phase (top $x$ axis). The numbers
identify a region that contributes to a peak in the function $P\left(
m_{z}\right) $ shown in Fig. 6(a)$.$ Here $a$ is the lattice spacing and $%
\ell $ is the magnetic length. The dashed curve in (b) shows the
magnetization trace for $N=3,M=2$ at $\protect\nu =0.25.$}
\label{figure5}
\end{figure}
\begin{figure*}[tbph]
{\includegraphics[scale=1]{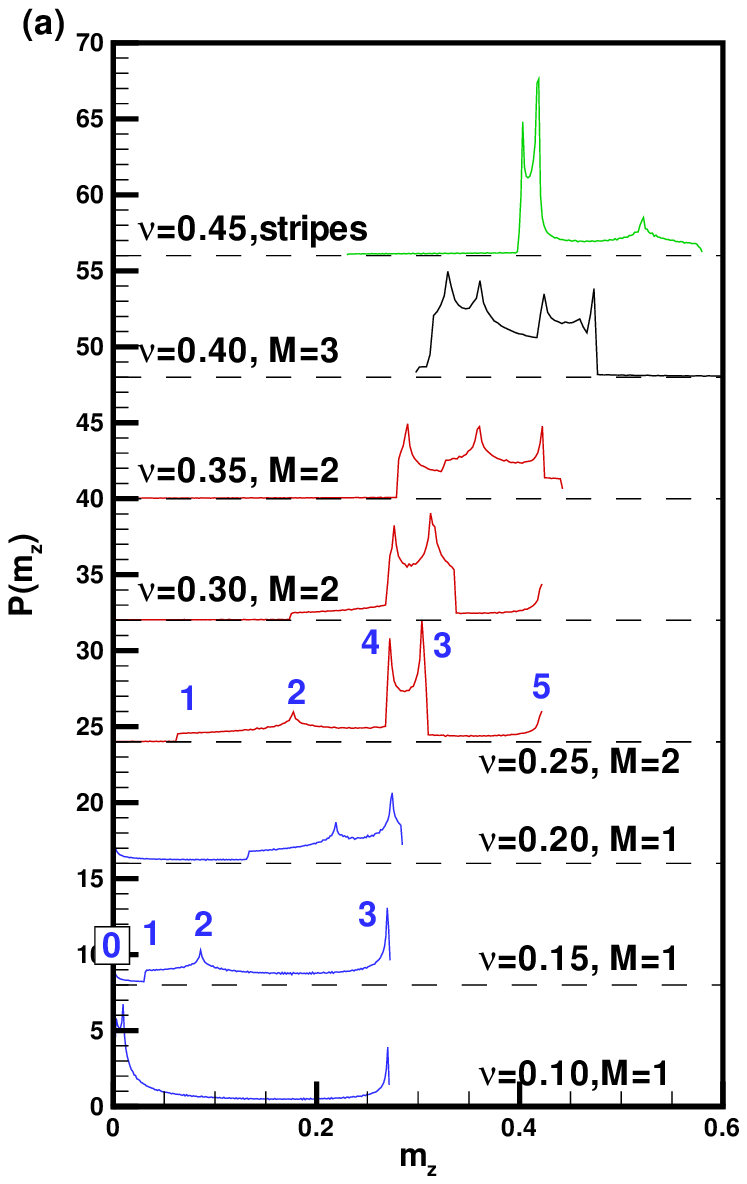}} \quad {%
\includegraphics[scale=1]{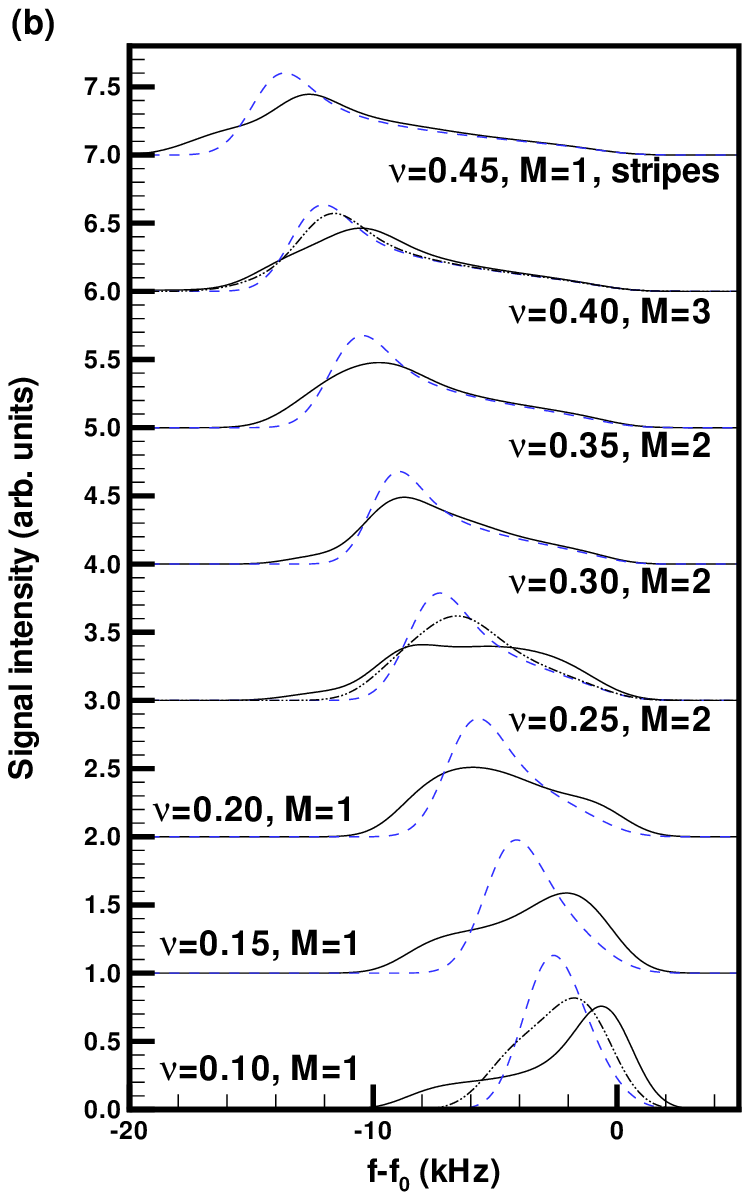}}
\caption{(Color online) (a) Probability density $P\left( m_{z}\right) $ for
some BC1 and SC1 phases in Landau level $N=2$ at different filling factors $%
\protect\nu <0.5.$ Each curve is offset by $8$ from the one below for
clarity. This offset must be deducted from the value of $P\left(
m_{z}\right) $; (b) corresponding RD-NMR spectra for the crystal phases in
(a) (full black curve) and for the uniform phases (dashed blue curves). Each
curve is offset by 1 from the one below for clarity. The intensity is in
arbitrary units, but the same for all curves. The dashed-dotted curves are
the spectra of the crystal phases with a blurring factor $\protect\sigma %
=1.4.$}
\end{figure*}
\begin{figure}[tbph]
\includegraphics[scale=0.8]{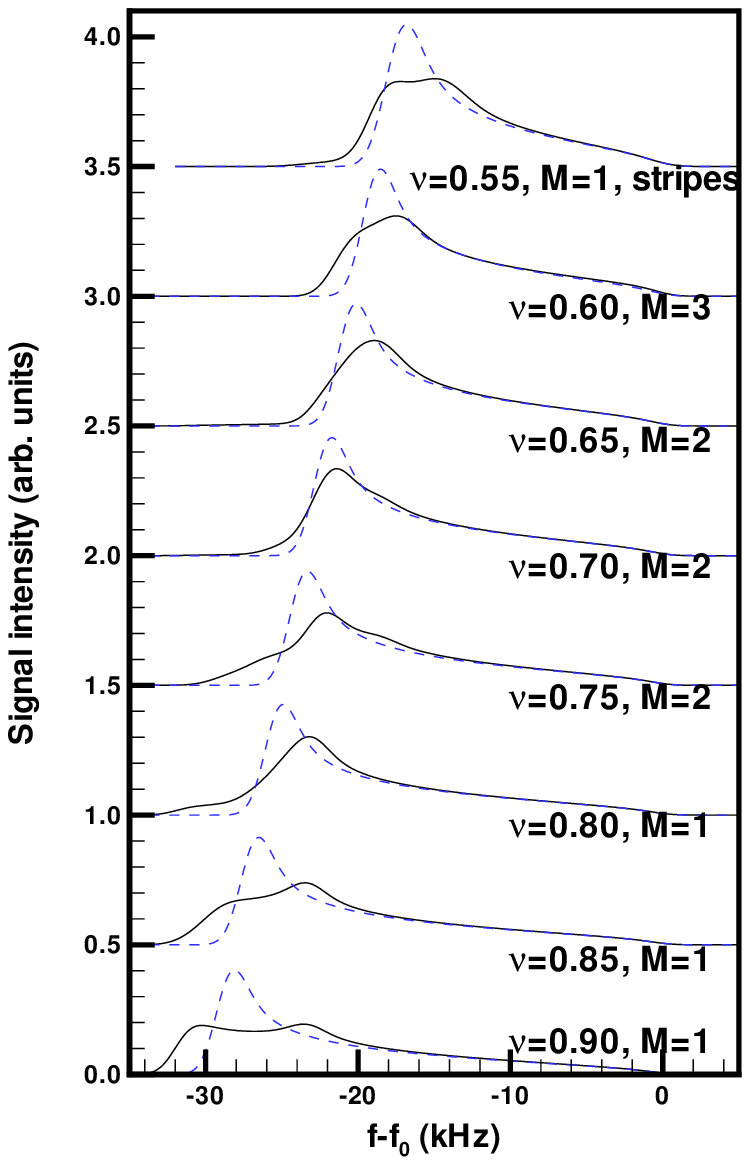}
\caption{(Color online) RD-NMR spectra for the BC2 crystal phase (full black
curve) and uniform phase (blue dashed curves) in Landau level $N=2$ for
filling factor $0.5<\protect\nu <1.$ The curves are offset by $0.5$ for
clarity. The intensity is in arbitrary units, the same for all curves. }
\end{figure}

We now consider Landau level $N=2.$ Figure 1(a) of Ref. %
\onlinecite{Cotebubble} shows the phase diagram of the 2DEG in Landau level $%
N=2$ for $0\leq \nu \leq 0.5.$ The density pattern for each phase (BC1 and
SC1) is shown in Fig. 2 of that reference. In the HFA, there is a transition
at $\nu =0.22$ from $M=1$ to $M=2$ and at $\nu =0.37$ a second transition to 
$M=3$. The stripe phase SC1 is reached after $\nu =0.43.$ These numbers are
for a perfect 2DEG. They are slightly modified when $W$ is finite as is the
case here. Figure 5 shows the polarization $m_{z}\left( \mathbf{r}\right) $
along $x$ for $y=0$ in the bubble and stripe phases at different values of
the filling factors$.$ The corresponding functions $P\left( m_{z}\right) $
and line shapes $I_{\nu }\left( f\right) $ are shown in Fig. 6(a) and Fig.
6(b) respectively. The bubble crystals and stripes of holes BC2, SC2 follow
the same phase diagram than BC1, SC1 but in reverse order, i.e., the crystal
of electrons at $\nu <0.5$ is replaced by a crystal of holes at filling
factor $1-\nu $ and $m_{z}$ is replaced by $1-m_{z}$ in the calculation of
the line shape. The spectra of the phases BC2, SC2 are shown in Fig. 7. The
density $m_{z}$ being bigger in the hole phases than in the electron phases,
the Knight shift is larger in the former than in the latter.

As $\nu $ increases, the radius of the ring in the $M=1$ bubble crystal
increases. The transition to the $M=2$ crystal occurs when rings from
adjacent sites just touch. At this point, there is an abrupt transition to
the $M=2$ bubble crystal which has two rings. As $\nu $ increases further,
these two rings expands and there is again an abrupt transition to the $M=3$
bubble crystal when the outer rings from adjacent sites just touch. The $M=3$
bubble crystal has two rings and one central maximum on each site. From Fig.
5, as $\nu $ increases, the average value of the magnetization increases and
there are more oscillations in $m_{z}\left( \mathbf{r}\right) $ and
consequently more structure in $P\left( m_{z}\right) .$ In each case, there
is one oscillation with a large value of $\Delta m_{z}\left( \mathbf{r}%
\right) $ and other smaller ones. In the stripe phase, the oscillations are
much less pronounced than in the crystal phases. As regards its NMR
spectrum, this phase is close to that of the uniform phase.

In Fig. 5, the numbers above the lines for $\nu =0.15$ and $\nu =0.25$
identify the regions in $m_{z}\left( \mathbf{r}\right) $ that are
responsible for a peak in the corresponding plot of $P\left( m_{z}\right) $
shown in Fig. 6(a). We see in Fig. 5(b) that, in comparison with $M=1,$ the $%
M=2$ bubble crystal has one extra ring which adds the extra peaks numbered $%
3 $ and $4$ in Fig. 6(a) for $\nu =0.25.$

The spectral line shape of the BC1, SC1 crystals changes with increasing $%
\nu $ and is different from that of the uniform phase. The signal from the $%
M=2$ crystal at $\nu =0.25$ looks different from that of the $M=1$ crystal
at $\nu =0.1$ but is not very different from the $M=1$ crystal at $\nu =0.2.$
Even though the functions $P\left( m_{z}\right) $ shown in Fig. 6(a) are
quite different, many of the peaks in $P\left( m_{z}\right) $ are too close
in $m_{z}$ to be resolved in the spectra. For example, the peaks numbered $3$
and $4$ in Fig. 6(a) (for $\nu =0.25,M=2$) are spaced by $\Delta
m_{z}=0.025. $ This leads to peaks separated by $\Delta f=0.8$ kHz in the
line shape. This separation is smaller than the linewidth $\Gamma =1.5$ kHz
considered in our calculation. It follows that most of the interesting
features in $P\left( m_{z}\right) $ are lost in the integrated signal $%
I_{\nu }\left( f\right) $ (not considering the extra complication due to
confining wave function). The signal from the $M=3$ BC1 crystal, however, is
different from that of the $M=1,2$ crystal and the stripe phase. It has a
characteristic long low frequency tail (not visible in Fig. 6(a) but it is
finite up to $m_{z}=1$ and in Fig. 6(b) it extends to $f-f_{0}=-32$ kHz). As
we mentioned above, this phase may not be present in the ground state
according to DMRG calculations. We expect the same conclusions to hold for
Landau levels $N>2$ since there are more oscillations in bubbles with higher
values of $M$ and these oscillations are smaller in amplitude [see the
traces for $M=2,\nu =0.25$ and $N=2,3$ in Fig. 5(b)]. Except in some range
of $\nu $ (fort $\nu =0.3-0.35,$ for example) most spectra are sufficiently
different from that of the corresponding uniform state to allow a
discrimination between the uniform and crystal states. Unfortunately, it
does not seem possible to infer the number of electrons in each bubble
crystal from its spectrum alone when $N\geq 2$ (with the exception of $M=N+1$%
) since the distinguishing features in the density are lost in the
integrated signal. It may also be difficult to differentiate between the $M=2
$ crystal and the stripe phase as their spectra are not dramatically
different. Moreover, the oscillations are small in the stripe phase and the
corresponding NMR spectrum is close to that of the uniform state. One marked
difference between the $M=2$ and $M=1$ spectra, however, is that the maximal
Knight shift (in absolute value) is larger in the former than in the latter.
Indeed, as Fig. 6(a) shows, the maximum value of $m_{z}\left( \mathbf{r}%
\right) $ increases discontinuously with $M.$

\begin{figure}[tbph]
\includegraphics[scale=0.8]{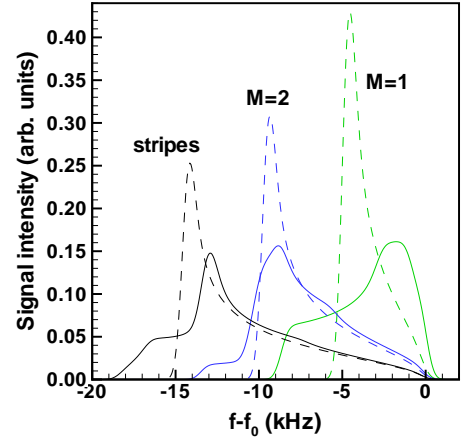}
\caption{(Color online) $N=2$ RD-NMR spectra (full curves) for the $M=1$ and 
$M=2$ bubble crystals at $\protect\nu =0.15$ and $\protect\nu =0.3$ and for
the stripe phase at $\protect\nu =0.45$ calculated with a linewidth $\Gamma
=0.5$ kHz. The dashed curves are the spectra of the corresponding uniform
phases.}
\end{figure}

At some filling factors in $N=2$, the difference between the crystal and
liquid spectra is small and may be washed out by fluctuations. We give some
examples of that in Fig. 6(b) where the dashed-dotted curve represent the
crystal phases calculated with a Gaussian blurring factor $\sigma =1.4$
which is the value necessary in $N=0$ to make the calculated spectrum agrees
with the experimental result according to Ref. \onlinecite{Tiemann}. We see
that, with the exception of $\nu =0.1,$ the spectra for $\nu =0.25$ and $\nu
=0.4$ are almost identical with those of the corresponding uniform states.
We checked that the value of $\sigma $ that brings the crystal spectrum
identical with that of the uniform state for the Wigner crystal at $\nu =0.1$
in $N=0,1,2$ are $\sigma =4.5,4.0,3.5$ respectively. The crystal state is
thus more fragile in higher Landau levels.

\section{CONCLUSION}

We have presented in this paper a study of the RD-NMR\ spectra of various
crystal phases that can theoretically exist as ground states of the 2DEG in
the quantum Hall regime. In Landau levels $N=0,1$, Wigner crystal, bubble
crystal, stripe phase and Skyrme crystal (when present) produce distinct
RD-NMR line shapes that should allow their unambiguous experimental
identification. In higher Landau levels, the RD-NMR spectra are not so
distinct with the exception of the $M=N+1$ bubble crystal. In these levels,
crystal and uniform states can be distinguished but it would be difficult to
differentiate between the Wigner, bubble (with $M>1$ and $M<N+1$) and stripe
phases.

Our calculation assumed the ideal case of a frozen solid at zero
temperature. As shown before\cite{Tiemann,Tiemann2}, thermal fluctuations
will render the crystal spectra closer to that of the uniform state. This
will not help the differentiation of the different phases in $N\geq 2$ since
the small difference in the spectra will probably be washed out unless the
temperature is very small.

The NMR\ spectra would be more discriminating with a stronger hyperfine
coupling between the electron gas and the nuclei or with a reduced linewidth 
$\Gamma $ but these parameters are fixed by the quantum well. We have used
the value $\Gamma =1.5$ kHz in all our calculations\cite{Tiemann} but, as a
test, we show in Fig. 8 the spectra for the $N=2$ stripe phase at $\nu =0.45$
and the $N=2,M=1,2$ bubble crystals at $\nu =0.15$ and $\nu =0.3$ with a
smaller linewidth $\Gamma =0.5$ kHz. The distinction between the solid and
uniform phases is much clearer [compare with Fig. 6(b)]. Apart from the
important difference in the range of the Knight shift for crystal with
different values of $M$, the spectra for the $M=2$ crystal and the stripe
phase still have a similar shape.

In a real experimental situation, the filling factor is tuned by changing
the potential on a gate in the GaAs/AlGaAs quantum well. This modifies the
shape of the confining wave function that enters the calculation of the NMR\
spectra. We have considered this wave function as fixed, but our calculation
can easily take this effect into account (by solving the Schr\"{o}%
dinger-Poisson equations) if an exact description of the quantum well is
given.

\begin{acknowledgments}
R. C. was supported by a grant from the Natural Sciences and Engineering
Research Council of Canada (NSERC). Computer time was provided by Calcul Qu%
\'{e}bec and Compute Canada.
\end{acknowledgments}

\end{document}